\documentclass[aps,prl,reprint,superscriptaddress,amsmath,amssymb,amsfonts]{revtex4-2}

\usepackage{graphicx}% Include figure files
\usepackage{dcolumn}% Align table columns on decimal point
\usepackage{bm}% bold math
\usepackage{hyperref} % add hypertext capabilities
\usepackage[mathlines]{lineno}% Enable numbering of text and display math

\usepackage{soul}
\usepackage{xcolor}
\definecolor{lightblue}{rgb}{0.5,0.8,1}
\sethlcolor{lightblue}
\soulregister\cite7
\soulregister\parencite7
\soulregister\textcite7
\soulregister\autocite7

\begin{document}

%\preprint{}

\title{Coherent regime of Kapitza-Dirac effect with electrons}


\author{Kamila Moriová}
    \affiliation{Faculty of Mathematics and Physics, Charles University, Ke Karlovu 3, 12116 Prague 2, Czech Republic}
\author{Petr Koutenský}
    \affiliation{Faculty of Mathematics and Physics, Charles University, Ke Karlovu 3, 12116 Prague 2, Czech Republic}
\author{Neli Laštovičková Streshkova}
    \affiliation{Faculty of Mathematics and Physics, Charles University, Ke Karlovu 3, 12116 Prague 2, Czech Republic}
\author{Marius Constantin Chirita Mihaila}
    \affiliation{Faculty of Mathematics and Physics, Charles University, Ke Karlovu 3, 12116 Prague 2, Czech Republic}
\author{Zbyněk Šobáň}
    \affiliation{Institute of Physics, Czech Academy of Sciences, Cukrovarnická 10, 162 00 Prague 6, Czech Republic}
\author{Jaromír Kopeček}
    \affiliation{Institute of Physics, Czech Academy of Sciences, Cukrovarnická 10, 162 00 Prague 6, Czech Republic}    
\author{Andreas Schertel}
    \affiliation{Carl Zeiss Microscopy GmbH, Carl-Zeiss-Straße 22, 73447 Oberkochen, Germany}
\author{Martin Kozák}
    \email{m.kozak@matfyz.cuni.cz}
\affiliation{Faculty of Mathematics and Physics, Charles University, Ke Karlovu 3, 12116 Prague 2, Czech Republic}
\thanks{Corresponding author: m.kozak@matfyz.cuni.cz}

\begin{abstract}
Electron matter waves coherently diffract when passing through a periodic structure of light formed by two interfering light waves. In this so-called Kapitza-Dirac effect, the electron momentum changes due to absorption and emission of photons via stimulated Compton scattering. Until now, the effect has only been observed with low energy electrons due to the small momentum of a visible photon compared to the momentum of high energy electron leading to diffraction angles of $10^{-4}$ rad or smaller. We report on the observation of the Kapitza-Dirac effect in a scanning electron microscope using high energy (20 keV and 30 keV) electrons with de-Broglie wavelengths of 9 pm and 7 pm, respectively. The photon sidebands in the electron transverse momentum spectrum are detected in the convergent beam diffraction geometry using spatial filtering. As the coupling strength between the electrons and the light field increases, the sideband populations exhibit coherent, reversible oscillations among diffraction orders. The effect can serve as a coherent electron beam-splitter or a phase-plate in various types of electron microscopes. 
\end{abstract}

\maketitle

The diffraction of coherent waves at periodic structures is one of the fundamental phenomena in physics. The effect can be understood as an interference of waves generated by the individual sources represented by neighboring periods of the structure that scatter the incident wave. In addition to diffraction of light waves at solid gratings, the diffraction of matter waves at a grating formed by light was proposed by Kapitza and Dirac \cite{Kapitza_Dirac1933}. In this process, the electron wave interacts with the ponderomotive potential of an optical standing wave, which is linearly proportional to the local light intensity. The light interference leads to the formation of a periodic potential, which imparts a phase modulation to the wave function of an electron propagating through the light grating. The phenomenon can also be viewed in the particle picture as stimulated Compton scattering, in which the electron absorbs and emits photons from the two light waves in a stimulated manner \cite{Batelaan2007}.

Kapitza-Dirac effect was first observed with atoms \cite{Gould1986,Giltner1995}, where the interaction can be enhanced by the presence of electronic resonances. The scattering of electron matter waves at optical fields has initially been demonstrated in the incoherent interaction regime, in which the individual diffraction orders are not resolved \cite{Bucksbaum1987,Bucksbaum1988}. In the incoherent regime, diffraction peaks are absent because the transverse coherence length of the electron beam is significantly shorter than the period of the optical standing wave. The incoherent electron scattering at optical standing waves generated by ultrashort laser pulses has been applied for characterization of electron pulses with sub-picosecond time resolution \cite{Hebeisen_2008,Gao_2012,Moriova_2025}. Only in 2001, the coherent diffraction peaks were observed \cite{Freimund2001} using electron beam with energy of 380 eV (de Broglie wavelength of $6.3 \times 10^{-11}$ m). The observation was enabled by spatial filtering of the electron beam to enhance its spatial coherence, together with detection providing high transverse momentum resolution. Since then, the use of ponderomotive potential of optical fields for modulating electron beams has been studied extensively both theoretically \cite{Li_2004,Smirnova_2004,Kaplan_2005,Smorenburg_2011,McGregor_2015,Erhard_2015,Dellweg2017,Abajo2021,Ebel_2025,Chahshouri_2026} and experimentally \cite{Axelrod_2020,ChiritaMihaila_2022,Lin_2024,ChiritaMihaila_2025}. The process has also been generalized to allow efficient modulation of the longitudinal momentum of electrons using two-color fields \cite{Baum_2007,Hilbert_2009,Kozák_2018,Kozak_2018b,Kozak_2019,Tsarev_2023} and the effects beyond non-recoil approximation were studied theoretically \cite{Talebi_2018, Kozak_2022}. The extended interest in Kapitza-Dirac effect is mainly due to its potential applications for coherent splitting of electron beams or for contrast enhancement due to the phase modulation induced in the electron wave \cite{Schwartz2019}.

Electron diffraction at optical gratings can occur in two distinct regimes depending on the effective length of the interaction. In the diffraction regime (referred to as the Raman-Nath regime) corresponding to the interaction of the electron wave with the focused light beams, the final electron state can be written as a superposition of plane waves corresponding to individual diffraction orders whose transverse momenta differ by $2\hbar k$. Here $k=2\pi/\lambda$ is the length of the wave vector of the light waves with wavelength $\lambda$, which form the optical grating. In the long-interaction regime (Bragg regime), a single diffraction peak is observed in the electron distribution when the Bragg condition for the incident electron wave is fulfilled \cite{Freimund_2002}.

Here we investigate the coherent Kapitza-Dirac effect in the Raman–Nath regime. We observe, that as the laser intensity is increased, the populations of the photon sidebands first grow and then exhibit coherent oscillations. These oscillations, driven by the increasing coupling strength between the electron and the optical field, reveal the reversible population exchange among diffraction orders. Such behavior has not been observed in previous Kapitza–Dirac experiments with electrons. Our measurements constitute the first realization of an elastic coherent electron–photon interaction experiment performed in a scanning electron microscope, opening new possibilities for time-resolved and light–matter–coupled electron microscopy.

We demonstrate the Kapitza-Dirac effect with fast electrons with a kinetic energy of 20 keV (de Broglie wavelength 8.5 pm) and 30 keV (7 pm) in an ultrafast scanning electron microscope with field emission electron source \cite{Moriova_2025} (details of the experimental setup are described in Supplemental Material). The layout of the setup is shown in Fig. 1. The optical standing wave is generated by two counterpropagating pulsed laser beams with a wavelength of $\lambda$=1030 nm (photon energy of 1.2 eV) and pulse durations of 220 fs in the first set of experiments and 700 fs in the second set of experiments demonstrating the coherent oscillations of diffraction probabilities, respectively, which are focused to the beam radius of approximately 10 $\mu$m. Polarization of both beams was linear in the direction of the electron beam propagation (along \textit{z} axis). The electron wave packets are generated in the form of short pulses that are photoemitted from the microscope Schottky tip using ultraviolet (photon energy of 4.8 eV) or visible (2.4 eV) pulses. The arrival times of both laser pulses with respect to the electron pulse are controlled with two independent optical delay lines to ensure time overlap in the interaction region. The spatial overlap is adjusted by using light scattering from the apex of a tungsten nanotip, which is aligned to the center of the electron column by secondary electron imaging. The nanotip is retracted for the experiments.

To resolve individual photon orders in the transverse momentum spectrum of the scattered electrons, the angular resolution of the detection setup has to be better than $10^{-5}$ rad. This is a much smaller value than the typical divergence angle of an electron beam in scanning electron microscopes. The angular resolution in diffraction experiments can be significantly improved in a convergent beam electron diffraction (CBED) geometry. When the optical standing wave is placed upstream the focus of an electron beam, the high-resolution diffraction pattern is obtained in the focal plane. For electrons with kinetic energy of 20 keV and the laser wavelength of 1030 nm, the angle between the two neighboring diffraction orders is $\delta=1.73 \times 10^{-5}$ rad. To obtain sufficient spatial resolution allowing us to observe the individual diffraction peaks, we use a nanoslit with the width of 70 nm, which is cut using focused ion beam into a 50 nm thick Si$_3$N$_4$ membrane coated with several nm of Pt. The nanoslit is placed in the focus of the electron beam in the working distance of 24 mm in the direction perpendicular to the wave vector of the optical standing wave. The nanoslit acts as a spatial filter transmitting only a 70 nm wide stripe of the electron beam. The optical grating is placed 12 mm upstream of the slit, which gives us the spatial separation between the individual diffraction peaks in the focal plane $\Delta x=$208 nm, which is about three times higher than our resolution given by the convolution of the nanoslit width and the transverse size of the focused electron beam of approximately 20 nm. The electron beam generating the diffraction pattern is scanned across the slit and the transmitted electron current is detected using the hybrid pixel detector (TimePix3). We note that elastic or inelastic scattering of electrons at the slit does not play any role because the only detected quantity is the total electron current as a function of the position of the beam focus with respect to the nanoslit.

\begin{figure}
\includegraphics{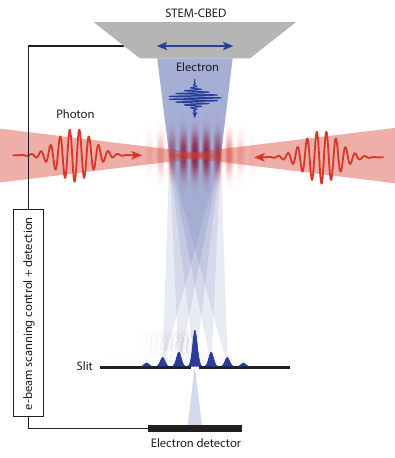}
\caption{(a) Layout of the experimental setup used for demonstration of Kapitza-Dirac effect with fast electrons based on scanning transmission electron microscopy combined with convergent beam electron diffraction (STEM-CBED). Pulsed convergent electron beam diffracts at an optical standing wave placed upstream the beam focus. A nanoslit with the width of 70 nm is placed to the electron beam focus to spatially filter a narrow distribution of electron transverse momentum. The electron diffraction pattern is measured by scanning the electron beam and detecting the electrons transmitted through the slit.}
\label{fig1}
\end{figure}

The interaction between an electron matter wave and an optical standing wave in vacuum can be described using semiclassical or fully quantum approaches \cite{Batelaan2007,Efremov1999}. When the standing wave is formed by coherent light associated with a classical oscillating electromagnetic field, we can apply the semiclassical treatment. When propagating through the optical fields, the electron wave function acquires a spatially dependent phase modulation owing to the interaction with the ponderomotive potential. In the weak interaction regime, in which the change of the electron momentum is negligible compared to its initial momentum (nonrecoil approximation), the phase acquired by the electron wave function can be expressed as:

\begin{equation}
\label{eq:01}
    \Phi(\mathbf{r},t)=-\frac{1}{\hbar}\int_{t_{0}}^{t} H_{\text{int}}\left(\mathbf{r}\left(z-vt+vt'\right),t'\right)\,\mathrm{d} t',
\end{equation}

where $H_\text{int}\left(\mathbf{r}\left(z-vt+vt'\right),t'\right)$ is the interaction Hamiltonian and $\mathbf{r}(z-vt+vt')$ is the classical unperturbed trajectory of the electron. The wave function of the electron immediately after the interaction can be written as $\psi(\mathbf{r},t)=e^{i\Phi(\mathbf{r},t)}\psi_\text{ini}(\mathbf{r},t)$. The Hamiltonian describing the interaction of a free charged particle with electromagnetic fields in nonrecoil approximation has a form \cite{Abajo2021}:

\begin{equation}
\label{eq:02}
    \begin{aligned}
    H_\text{int}\left(\mathbf{r},t\right)=\frac{e{\mathbf{p}}.\mathbf{A}\left(\mathbf{r},t\right)}{m_e}-e\varphi\left(\mathbf{r},t\right)+\\
    +\frac{e^2}{2m_e\gamma}\left [ A_x^2\left(\mathbf{r},t\right)+A_y^2\left(\mathbf{r},t\right)+\frac{A_z^2\left(\mathbf{r},t\right)}{\gamma^2}\right].
    \end{aligned}
\end{equation}

Here $e$ and $m_e$ are electron charge and mass, respectively, $\mathbf{p}$ is the initial electron momentum, $\mathbf{A}\left(\mathbf{r},t\right)$ is the vector potential of the electromagnetic field, $\varphi\left(\mathbf{r},t\right)$ is the scalar potential and $\gamma=1/\sqrt{1-v^2/c^2}$ is the Lorentz factor of the electron propagating with velocity $v$. We assume a gauge in which $\varphi\left(\mathbf{r},t\right)=0$, $E\left( \mathbf{r},t\right)=-\frac{\partial A\left( \mathbf{r},t\right)}{\partial t}$ and Coulomb gauge ($\nabla . \mathbf{A}\left( \mathbf{r},t\right)=0$). The electron propagates initially along the $z$-coordinate with the velocity $v_z$, which does not change during the interaction with the optical standing wave. Because the interaction occurs in vacuum and the interaction time is much longer than the oscillation period of the optical wave, the $\mathbf{p}.\mathbf{A}$ term in Eq. (\ref{eq:01}) vanishes after integration over time of the interaction \cite{Abajo2021} and the only nonzero contribution is given by the ponderomotive potential, that is, quadratic function of the vector potential. The vector potential of the optical standing wave with the electric field polarized along the $z$ direction, which is generated as a superposition of two counter-propagating waves with vector potentials $A_{\pm}\left(\mathbf{r},t\right)=\frac{A_0}{2}\cos(\omega t\pm kx)g_{\pm}\left(\mathbf{r},t\right)$, is $A_z\left(\mathbf{r},t\right)=A_0\cos(kx)\cos(\omega t)g\left(\mathbf{r},t\right)$. Here $\omega$ is the light frequency and $g\left(\mathbf{r},t\right)=g_{+}\left(\mathbf{r},t\right)+g_{-}\left(\mathbf{r},t\right)$ is a slowly-varying spatio-temporal envelope of the optical standing wave given by $g_{\pm}\left(\mathbf{r},t\right)=\exp{\left(-\frac{y^2+z^2}{w_0^2} \right)} \exp{\left[  -4\ln{2}  \frac{\left( t \mp x/c \right)^2}{\tau_l^2}\right]}$. Here $w_0$ and $\tau_l$ denote the beam waist and the laser pulse duration, respectively. When we introduce the vector potential of the optical standing wave to Eqs. (\ref{eq:01}) and (\ref{eq:02}), the final phase-modulated electron wave function can be written as a superposition of discrete momentum states (Jacobi-Anger expansion) as:

\begin{equation}
\label{eq:03}
    \begin{aligned}
    \psi(\mathbf{r},t)=\psi_{ini}(\mathbf{r},t)\sum_{n=-\infty}^{\infty}i^n J_n(\beta(\mathbf{r},\tau))e^{i2nkx}.
    \end{aligned}
\end{equation}

The final state thus can be viewed as a superposition of plane waves with transverse momenta shifted by $\pm 2\hbar k$, whose amplitudes are proportional to the Bessel functions of the $n$-th order $J_n(\beta(\mathbf{r},\tau))$ of the argument:

\begin{equation}
\label{eq:04}
    \begin{aligned}
    \beta\left( x,y,\tau\right)=-\frac{e^2 E_0^2}{8 m_e \hbar \gamma^3 \omega^2 v_z}\int_{-z_{\text{int}}}^{z_{\text{int}}} g\left(\mathbf{r'},\frac{z'}{v_z}-\tau\right) dz',
    \end{aligned}
\end{equation}

where $z_{\text{int}}$ is the distance at which the envelope of the laser fields decreases to zero. When we assume a Gaussian envelope of the optical fields in both space and time, condition $z_{\text{int}}=\infty$ can be used. In experiments, the finite envelopes of the light fields and electron pulses influence the measured populations of individual states in the transverse momentum space due to the dependence of the coupling parameter $\beta$ on the spatial coordinates $x$ and $y$ and the time delay of the electron wave with respect to the pulsed optical standing wave $\tau$. The population in each transverse momentum state is given by the integration over the initial electron distribution.

In the particle picture, the interaction can be described as a stimulated scattering, in which the electron absorbs and emits photons from the two coherent optical waves. The electron makes transitions between the individual momentum states, whose quantum mechanical amplitudes may interfere depending on the experimental conditions. When the transverse size of the electron beam is comparable or larger than the optical standing wave and/or the duration of the electron pulse is longer than the duration of the optical pulses, the coupling parameter $\beta$ has a broad distribution, which after integration over the interaction leads to suppression of interference. In this regime, the observed diffraction patterns broaden with increasing intensity of the optical standing wave, but no oscillations are observed in the populations of individual states. When we reach the regime in which all electrons experience approximately the same value of the coupling parameter $\beta$, the populations are expected to oscillate as a function of the interaction strength similarly to the case of the longitudinal modulation demonstrated in \cite{Feist2015}. In the case of coherent optical field generating the optical standing wave, the particle picture and the semiclassical picture are fully equivalent.

We experimentally demonstrate the Kapitza-Dirac type scattering in two regimes. In the experiments, we clearly resolve the individual diffraction orders corresponding to the integer number of transverse momentum $\pm 2n\hbar k$ absorbed by the electron during each stimulated absorption and emission of two photons. When the duration of the electron pulses is longer than the duration of the laser pulses forming the optical standing wave, we observe a broadening of the scattering patterns with increasing power (shown in Fig. 2a, b). Due to the broad distribution of the coupling parameter experienced by the electrons, there are no coherent oscillations observed in the populations of the individual photon sidebands. The experimental data measured with 20 keV electron energy and laser pulse duration of 220 fs are compared with the theoretically calculated scattering patterns (dashed curves) in Fig. 2b. Theoretical curves were calculated for $\beta_{\text{max}}=0, 1.8, 3.6, 5.4, 7.2 \text{ and }9$, where $\beta_{\text{max}}$ denotes the absolute value of the maximum coupling constant $\beta$. The spatial and temporal distribution of the coupling constant have been taken into account by integrating $\beta(x,z,\tau)$ in space and time with the weight corresponding to the spatiotemporal envelope of the electron pulse. Multiple photon orders are observed because the experiment occurs in the Raman-Nath (diffraction) regime of interaction. 

\begin{figure*}
\includegraphics[width=1\textwidth]{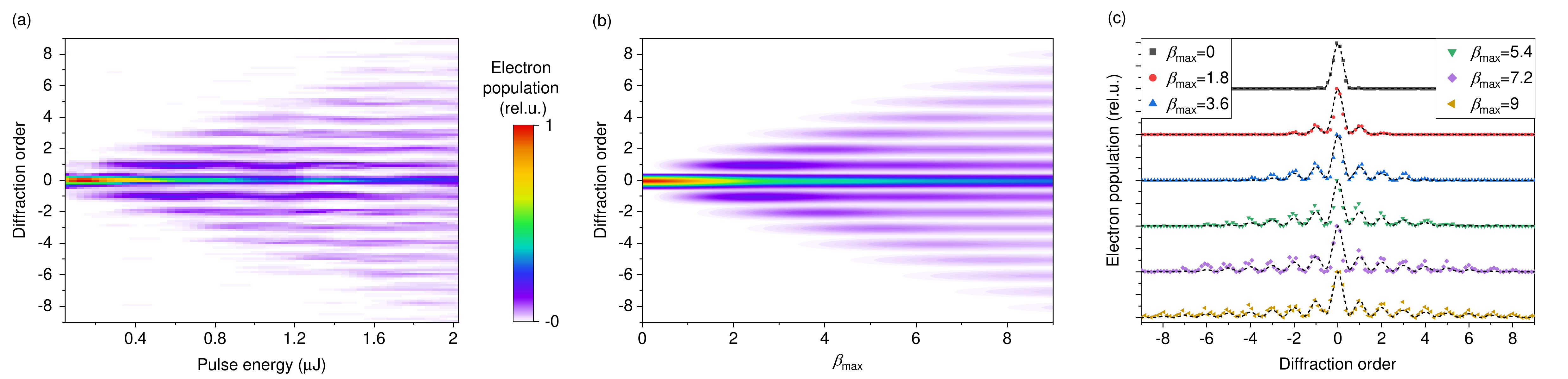}
\caption{Raman-Nath (diffraction) regime of Kapitza-Dirac effect with electron kinetic energy of 20 keV. (a) Measured electron transverse scattering patterns as a function of the laser pulse energy and (b) corresponding numerical calculations as a function of the maximum coupling constant $\beta_{\text{max}}$. (c) Normalized measured electron scattering patterns for selected values of $\beta_{\text{max}}$ (points) compared with the theory (dashed curves). Data are vertically shifted for clarity.}
\label{fig2}
\end{figure*}

To demonstrate the coherent regime of the interaction we increase the pulse duration of the laser pulses forming the optical standing wave to approximately 700-800 fs by adding negative second order dispersion using the compressor of the laser system. The visible pulses used for the photoemission of electrons in the electron gun of the microscope are recompressed by adding 20 cm of SF11 glass to the beam path. In this experiment we use the electron energy of 30 keV to ensure the shortest possible duration of the electron pulses of about 400 fs \cite{Moriova_2025}. Besides the condition that the electron pulse has to be shorter than the laser pulse, we also have to fulfill the condition of having the electron beam diameter significantly smaller than the spatial dimensions of the laser beams. However, in order to be able to observe diffraction peaks in the electron distribution in the focus of the beam, the dimension of the electron beam in the plane of interaction with the optical standing wave has to be larger than the spatial period of the ponderomotive potential. By selecting a smaller objective lens aperture, we limit the size of the electron beam in the interaction plane to approximately 3 $\mu$m. The beam convergence angle decreases to only $\approx$200 $\mu$rad.

\begin{figure*}
\includegraphics[width=1\textwidth]{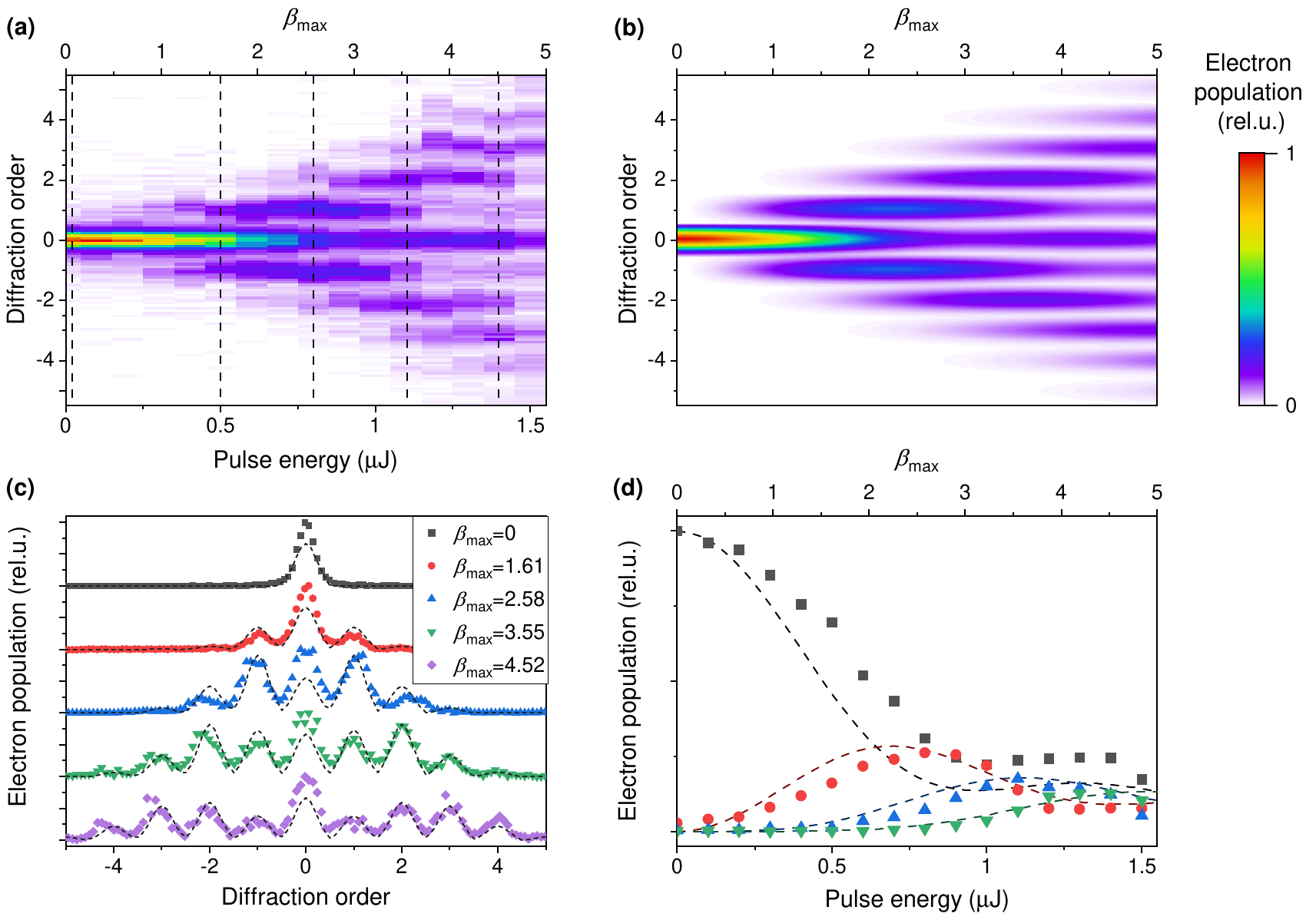}
\caption{Coherent Raman-Nath (diffraction) regime of Kapitza-Dirac effect measured with electron kinetic energy of 30 keV.(a) Measured and (b) calculated electron transverse scattering patterns as a function of the average laser power and coupling constant $\beta_{\text{max}}$. (c) Normalized measured electron scattering patterns for selected values of $\beta_{\text{max}}$ corresponding to the dashed lines in (a) (points) compared to theoretical simulations (dashed curves). Data are vertically shifted for clarity. (d) Measured populations of the individual diffraction orders (black squares corresponds to sideband $n=1$, red circles $n=2$, blue triangles $n=3$ and green triangles $n=4$) as a function of the laser power and coupling constant $\beta_{\text{max}}$ compared to the theoretical data (dashed curves). The population of each peak with $n\neq0$ is calculated as an average of populations in peaks $+n$ and $-n$.}
\label{fig3}
\end{figure*}

The data measured in the coherent interaction regime are shown in Fig. 3a. We observe that after the initial increase, the populations of the individual diffraction peaks with $n\neq0$ start to oscillate. These coherent oscillations have not been observed in previous Kapitza-Dirac experiments with electrons \cite{Freimund2001} but were observed in the Kapitza-Dirac scattering of atoms \cite{Gould1986}. The weaker interaction strength for the same pulse energy compared to Fig. 2a arises from the different pulse durations used in the two measurement sets. The experimental data are compared with the numerical simulations shown in Fig. 3b. In Fig. 3c we plot the measured (points) and calculated (dashed curves) electron scattering patterns for different values of the maximum coupling constant $\beta_{\text{max}}=0, 1.61, 2.58, 3.55 \text{ and }4.52$. The coherent oscillations of the populations of individual diffraction peaks are shown in Fig. 3d (points correspond to experimental data, dashed curves to numerical calculations).

We show that it is possible to observe the coherent diffraction regime of electron scattering with an optical standing wave in vacuum, the so-called Kapitze-Dirac effect, even with fast electrons with an extreme ratio of the de Broglie wavelength to the period of the optical standing wave of $1.73 \times 10^{-5}$. The CBED configuration can serve as a coherent beam splitter for electrons in electron microscopes, where it can generate several coherent beamlets in the focal plane of the electron beam. The coherent superposition can be used, for example, for sensitive interferometric experiments performed on the nanoscale (see \cite{SM} for details). When we assume that the quantum mechanical phase acquired by one of these beamlets changes due to the local interaction with the sample, the interference in the far-field will lead to spatial distortions or change of the intensity of the electron beam distribution, which can be detected by a position-resolving detector. Alternatively, the coherent beam splitting using Kapitza-Dirac effect may find application in ultrafast implementation of interaction free imaging \cite{Turner_2021}. The coherent interaction regime demonstrated here allows us to control the populations of individual photon sidebands by adjusting the coupling constant via control of the optical standing wave intensity. This type of control enables the opening and closing of channels, which can be utilized in multibeam quantum control schemes \cite{DiGiulio_2025}.

\begin{acknowledgments}
\paragraph{Acknowledgments}
Czech Science Foundation (project GA22-13001K), Charles University (UNCE/SCI/010, SVV-2020-260590, PRIMUS/19/SCI/05, GAUK 216222). Funded by the European Union (ERC, eWaveShaper, 101039339). This work was supported by TERAFIT project No. CZ.02.01.01/00/22\_008/0004594 funded by OP JAK, call Excellent Research.

\paragraph{Data Availability}
The data supporting the findings of
this study are openly available at \cite{data}.

\end{acknowledgments}

\nocite{*}

\bibliography{articlereferences}

\end{document}